\documentclass[11pt,a4paper]{article}
\usepackage{epsfig}
\begin{document}
\begin{flushright}
PRA-HEP 98-08
\end{flushright}
\begin{center}
{\Large \bf Factorization scheme analysis of $F_2^{\gamma}(x,Q^2)$
and parton distributions functions of the photon}
\footnote{Supported by the Grant Agency of ASCR under grant No. A1010602}

\vspace*{0.7cm}
Ji\v{r}\'{\i} Ch\'{y}la

\vspace*{0.2cm}
{\em Institute of Physics, Na Slovance 2, Prague 8, Czech Republic}

\vspace*{1cm}
\begin{quotation}
\noindent
Factorization scheme analysis of $F_2^{\gamma}(x,Q^2)$ in
the next--to--leading order QCD is revisited. It is emphasized that
the presence of the inhomogeneous term in the evolution equations
for quark distribution functions of the photon implies subtle but
important difference in the way factorization mechanism works in
photon--hadron and photon--photon collisions as compared to the
hadronic ones. It is argued that none of the existing NLO analyses
of $F_2^{\gamma}(x,Q^2)$ takes this difference properly into
account. The source of the ensuing incompleteness is traced back to
the misinterpretation of the behaviour of $q^{\gamma}(x,M)$ as a
function of $\alpha_s(M)$. Parton model interpretation of the so
called ``constant terms'' in the LO photonic coefficient function
$C_{\gamma}^{(0)}(x)$ is given and smooth transition between the
properties of virtual and real photon analyzed. Finally
phenomenological consequences of this analysis are discussed.
\end{quotation}

\end{center}

\section{Introduction}
Observed from a large distance the photon behaves as a neutral
structureless object governed by the laws of Quantum
Electrodynamics. However, when probed at short distances it
exhibits also some properties characteristic of hadrons. This
``photon structure'' is quantified, similarly as in the case of
hadrons, in terms of parton distribution functions (PDF),
satisfying certain evolution equations. Because of the direct
coupling photons to quark--antiquark pairs these evolution
equations are, contrary the case of hadrons, inhomogeneous. This
inhomogeneity has important implications for the way factorization
of mass singularities operates in collisions involving photons,
implications that have not been properly taken into account in
existing NLO analyses of $F_2^{\gamma}(x,Q^2)$. The primary aim of
this paper is to remove this shortcoming.

Secondly, we shall address several issues concerning the structure
of the virtual photon: transition between the properties of real
and virtual photon, properties and role of the longitudinal virtual
photon, and parton model interpretation of the so called ``constant
terms'' in LO photonic coefficient function $C_{\gamma}^{(0)}(x)$.

The paper is organized as follows. In the next Section basic facts
and notation concerning PDF of the photon are reviewed and the
properties of the pointlike part of quark distribution function of
the photon critically reanalyzed. In Section 3, which contains the
main result of this paper, we discuss in detail the factorization
scale and scheme dependence of $F_2^{\gamma}(x,Q^2)$ at the NLO and
point out the ingredients that must be included to make this
analysis complete. The properties of the virtual photon and the
transition of PDF of the virtual photon to those of the real one
are analyzed in Section 4. Phenomenological implications of the
present analysis are discussed in Section 5.

\section{Structure of the real photon}
Despite the recent progress in investigation of the structure of
the photon
\footnote{For recent theoretical and experimental reviews see
\cite{Vogt} and \cite{Stefan}, respectively.} our knowledge of the
properties of the photon still lags behind that of the nucleon. We
shall be primarily interested in strong interaction effects, but as
the basic ideas and formalism of the partonic structure of the
photon have a close analogy in QED, the latter will serve as a
guide in some of the following considerations.

\subsection{Notation and basic facts}
In QCD the coupling of quarks and gluons is characterized by the
renormalized colour coupling (``couplant'' for short)
$\alpha_s(\mu)$, depending on the {\em renormalization scale} $\mu$
and satisfying the equation
\begin{equation}
\frac{{\mathrm d}\alpha_s(\mu)}{{\mathrm d}\ln \mu^2}\equiv
\beta(\alpha_s(\mu))=
-\frac{\beta_0}{4\pi}\alpha_s^2(\mu)-
\frac{\beta_1}{16\pi^2}
\alpha_s^3(\mu)+\cdots,
\label{RG}
\end{equation}
where, in QCD with $n_f$ massless quark flavours, the first two
coefficients, $\beta_0=11-2n_f/3$ and $\beta_1=102-38n_f/3$, are
unique, while all the higher order ones are ambiguous. As we shall
stay in this paper within the NLO, only the first two, {\em
unique}, terms in (\ref{RG}) will be taken into account in the
following. Nevertheless, even for a given r.h.s. of (\ref{RG}) its
solution $\alpha_s(\mu)$ is not a unique function of $\mu$, because
there is an infinite number of solutions of (\ref{RG}), differing
by the initial condition. This so called {\em renormalization
scheme} (RS) ambiguity
\footnote{In higher orders this ambiguity includes also the
arbitrariness of the coefficients $\beta_i,i\ge 2$ in (\ref{RG}).}
can be parameterized in a number of ways. One of them makes use of
the fact that in the process of renormalization another dimensional
parameter, denoted usually $\Lambda$, inevitably appears in the
theory. This parameter depends on the RS and at the NLO actually
fully specifies it: RS=\{$\Lambda_{\mathrm {RS}}$\}. For instance,
$\alpha_s(\mu)$ in the familiar MS and $\overline{\mathrm {MS}}$ RS
are solutions of the same equation (\ref{RG}), but are associated
with different $\Lambda_{\mathrm {RS}}$
\footnote{The variation of both the renormalization scale $\mu$
and the renormalization scheme RS$\equiv$\{$\Lambda_{\mathrm {RS}}$\}
is actually redundant. It suffices to fix one of them and vary the
other, but I will stick to the common habit of considering both of
them as free parameters.}. In this paper we shall work in the
standard $\overline{\mathrm {MS}}$ RS of the couplant.

In QCD ``dressed'' PDF
\footnote{In the following the adjective ``dressed'' will be
dropped.} result from the resummation of multiple parton emissions
off the corresponding ``bare'' parton distributions. As a result of
this resummation PDF acquire dependence on the {\em factorization
scale} $M$. In parton model this scale defines the upper limit on
some measure $t$ of the off--shellness of partons included in the
definition of $D(x,M)$
\begin{equation}
D_i(x,M)\equiv \int_{t_{\mathrm {min}}}^{M^2}{\mathrm d}t
d_i(x,t),~~~~~~i=q,\overline{q},G,
\label{dressed}
\end{equation}
where the {\em unintegrated} PDF $d_i(x,t)$ describe distribution
functions of partons with the momentum fraction $x$ and {\em fixed}
off--shellness $t$. Parton virtuality $\tau\equiv\mid p^2-m^2\mid$
or transverse mass $m_T^2\equiv p_T^2+m^2$, are two standard
choices of such a measure. Because at small $t$, $d_i(x,t)={\cal
O}(1/t^k), k=1,2$, the dominant part of the integral
(\ref{dressed}) comes from the region of small off--shellness $t$.
Varying the upper bound $M^2$ in (\ref{dressed}) has therefore only
a small effect on the integral (\ref{dressed}), leading to weak (at
most logarithmic) scaling violations.

The factorization scale dependence of PDF of the photon
\footnote{If not stated otherwise all distribution functions
in the following concern the photon.} is determined by a system of
coupled inhomogeneous evolution equations
\begin{eqnarray}
\frac{{\mathrm d}\Sigma(x,M)}{{\mathrm d}\ln M^2}& =&
k_q+P_{qq}\otimes \Sigma+ P_{qG}\otimes G,
\label{Sigmaevolution} \\
\frac{{\mathrm d}G(x,M)}{{\mathrm d}\ln M^2} & =&
k_G+ P_{Gq}\otimes \Sigma+ P_{GG}\otimes G,
\label{Gevolution} \\
\frac{{\mathrm d}q_{\mathrm {NS}}^i(x,M)}{{\mathrm d}\ln M^2}& =&
\sigma_{\mathrm {NS}}^i k_q+P_{\mathrm {NS}}\otimes q_{\mathrm{NS}}^i,
\label{NSevolution}
\end{eqnarray}
where $\sigma^i_{\mathrm NS}\equiv (e_i^2/\langle
\!e^2\!\rangle-1)/n_f$ and the convolution $\otimes$ is defined
in a standard way as
\begin{equation}
P\otimes q(x)\equiv \int_x^1 \frac{{\mathrm d}y}{y}P(x/y)q(y).
\label{convolution}
\end{equation}
The singlet and nonsinglet quark distribution functions $\Sigma$
and $q^{\mathrm {NS}}_i$ are given as
\begin{eqnarray}
\Sigma(x,M) & \equiv &\sum_{i=1}^{n_f}
\left[q_i(x,M^2)+\overline{q}_i(x,M^2)\right]\equiv
\sum_{i=1}^{n_f}q_i^{+}(x,M^2),
\label{singlet} \\
q_{\mathrm {NS}}^i(x,M) & \equiv &q^+_i(x,M^2)-\Sigma(x,M)/n_f.
 \label{nonsinglet}
\end{eqnarray}
To order $\alpha$ the splitting functions $P_{ij}(x,M)$ and
$k_i(x,M)$ are given as power expansions in $\alpha_s(M)$:
\begin{eqnarray}
k_q(x,M) & = & \frac{\alpha}{2\pi}\left[k^{(0)}_q(x)+
\frac{\alpha_s(M)}{2\pi}k_q^{(1)}(x)+
\left(\frac{\alpha_s(M)}{\pi}\right)^2k^{(2)}_q(x)+\cdots\right],
\label{splitquark} \\
k_G(x,M) & = & \frac{\alpha}{2\pi}\left[~~~~~~~~~~~~
\frac{\alpha_s(M)}{2\pi}k_G^{(1)}(x)+
\left(\frac{\alpha_s(M)}{\pi}\right)^2k^{(2)}_G(x)+\cdots\right],
\label{splitgluon} \\
P_{ij}(x,M) & = &
~~~~~~~~~~~~~~~~~~\frac{\alpha_s(M)}{2\pi}P^{(0)}_{ij}(x) +
\left(\frac{\alpha_s(M)}{2\pi}\right)^2 P_{ij}^{(1)}(x)+\cdots,
\label{splitpij}
\end{eqnarray}
where the leading order splitting functions
$k_q^{(0)}(x)=3e_q^2(x^2+(1-x)^2)$ and $P^{(0)}_{ij}(x)$ are {\em
unique}, while all higher order ones
$k^{(j)}_q,k^{(j)}_G,P^{(j)}_{kl},j\ge 1$ depend on the choice of
the {\em factorization scheme} (FS)
\footnote{We can turn this statement around and consider
any factorization scheme FS to be specified by the corresponding
set of functions $k^{(j)}_q,k^{(j)}_G,P^{(j)}_{kl},j\ge 1$.}. The
photon structure function $F_2^{\gamma}(x,Q^2)$, measured in deep
inelastic scattering experiments on the (slightly off--shell)
photons \cite{PETRA,PEP,TRISTAN,LEP}, is given as the convolution
\footnote{The factor 2 accounts for the inclusion of antiquarks
in the sum.}
\begin{equation}
F_2^{\gamma}(x,Q^2)=\sum_{q}2x e_q^2
\left[q(M)\otimes C_q(Q/M)
+ G(M)\otimes C_G(Q/M)+C_{\gamma}^q(Q/M)\right]
\label{f2gamma}
\end{equation}
of photonic PDF and coefficient functions
$C_q(x),C_G(x),C^q_{\gamma}(x)$ admitting perturbative expansions
\begin{eqnarray}
C_q(x,Q/M) & = & \delta(1-x)~~~~~~~~~~~~~+
~~~~~~~~\frac{\alpha_s(\mu)}{2\pi}C^{(1)}_q(x, Q/M)+\cdots,
\label{cq} \\
C_G(x,Q/M) & = & ~~~~~~~~~~~~~~~~~~~~~~~~~~~~~~~~~~~
\frac{\alpha_s(\mu)}{2\pi}C^{(1)}_G(x,Q/M)+\cdots,
\label{cG} \\
C^q_{\gamma}(x,Q/M) & = &
e_q^2\frac{\alpha}{2\pi}C_{\gamma}^{(0)}(x,Q/M)+
e_q^2\frac{\alpha}{2\pi}
\frac{\alpha_s(\mu)}{2\pi}C_{\gamma}^{(1)}(x,Q/M)+\cdots.
\label{cg}
\end{eqnarray}
The renormalization scale $\mu$, used as argument of the expansion
parameter $\alpha_s(\mu)$, is in principle independent of the
factorization scale $M$. Note that despite the presence of $\mu$ as
argument of $\alpha_s(\mu)$ in (\ref{cq}--\ref{cg}), the
coefficient functions $C_q,C_G$ and $C_{\gamma}^q$ are actually
independent of $\mu$ because the $\mu$--dependence of
$\alpha_s(\mu)$ is cancelled by explicit dependence of $C^{(i)}_q,
C^{(i)}_G,C^{(i)}_{\gamma},i\ge 2$ on $\mu$ \cite{politzer}. On the
other hand, PDF and the coefficient functions $C_q, C_G$ and
$C^q_{\gamma}$ do depend on both the factorization scale $M$ and
factorization scheme FS=$\{k^{(i)}_q,k^{(i)}_G, P^{(i)}_{kl},i\ge
1\}$, but in such a correlated manner that physical quantities,
like $F_2^{\gamma}$, are independent of both $M$ and the FS,
provided expansions (\ref{splitquark}--\ref{splitpij}) and
(\ref{cq}--\ref{cg}) are taken to all orders in $\alpha_s(M)$ and
$\alpha_s(\mu)$. In practical calculations based on truncated forms
of (\ref{splitquark}--\ref{splitpij}) and (\ref{cq}--\ref{cg}) this
invariance is, however, lost and the choice of both $M$ and FS
makes numerical difference even for physical quantities. At the NLO
RS=\{$\Lambda_{\mathrm {RS}}$\} and
FS=\{$k_i^{(1)},k_i^{(2)},P^{(1)}_{ij}$\}. The expressions for
$C_q^{(1)},C^{(1)}_G$ given in \cite{bardeen} are usually claimed
to correspond to ``$\overline{\mathrm {MS}}$ factorization scheme''.
As argued in \cite{jch2}, this denomination is, however,
incomplete. The adjective ``$\overline{\mathrm {MS}}$'' concerns
exclusively the choice of the RS of the couplant $\alpha_s$ and has
nothing to do with the choice of the splitting functions
$P^{(1)}_{ij}$. The choices of the
renormalization scheme of the couplant $\alpha_s(M)$ and of the
factorization scheme of PDF are two completely independent
decisions, concerning two different and in general unrelated
redefinition procedures. Both are necessary in order to specify
uniquely the results of fixed order perturbative calculations, but
we may combine any choice of the RS of the couplant with any choice
of the FS of PDF. Note that the coefficient functions
$C_q^{(1)},C_G^{(1)},C^{(1)}_{\gamma}$ depend on both of them,
while the splitting functions depend only on the FS of PDF. The
results given in \cite{bardeen} correspond to $\overline{\mathrm
{MS}}$ RS of the couplant but to the ``minimal subtraction'' FS of
PDF
\footnote{See Section 2.6 of \cite{FP}, in particular eq. (2.31),
for discussion of this point.}. It is therefore more appropriate to
call this full specification of the renormalization and factorization
schemes as ``$\overline{\mathrm {MS}}+{\mathrm {MS}}$ scheme''.
Although the phenomenological relevance of treating $\mu$ and $M$
as independent parameters has been demonstrated \cite{fontannaz},
we shall follow the usual practice and set $\mu=M$.

\subsection{Properties of the pointlike solutions}
The general solution of the evolution equations
(\ref{Sigmaevolution}--\ref{NSevolution}) for the quark
distribution functions $q(x,M)$ can be written as the sum of a
particular solution of the full inhomogeneous equation, called {\em
pointlike} part, and the general solution of the corresponding
homogeneous equation, called {\em hadronic} part:
\begin{equation}
q(x,M)= q^{\mathrm {PL}}(x,M)+q^{\mathrm {HAD}}(x,M).
\label{separation}
\end{equation}
Having separated $q(x,M)$ into its hadronic and pointlike parts
we can insert it into the r.h.s. of the evolution equation
(\ref{Gevolution}) for the gluon distribution function and write
its solutions as a sum of the hadronic and pointlike parts. At the
LO, where $k_G=0$, both of these components satisfy standard
homogeneous evolution equation
\begin{equation}
\frac{{\mathrm d}G^k(x,M)}{{\mathrm d}\ln M^2}=
P_{Gq}\otimes \Sigma^k+
P_{GG}\otimes G^k,~~~k={\scriptstyle \mathrm {HAD,~PL}}
\label{Geq}
\end{equation}
However, as there is an infinite number of pointlike solutions
$q^{\mathrm PL}(x,M)$, which differ by terms satisfying the
homogeneous evolution equation, the above separation of the quark
and gluons distribution functions into their pointlike and hadronic
parts is not unique and consequently these concepts have
separately no physical meaning.
\begin{figure}\unitlength 1mm
\begin{picture}(150,60)
\put(0,30){\epsfig{file=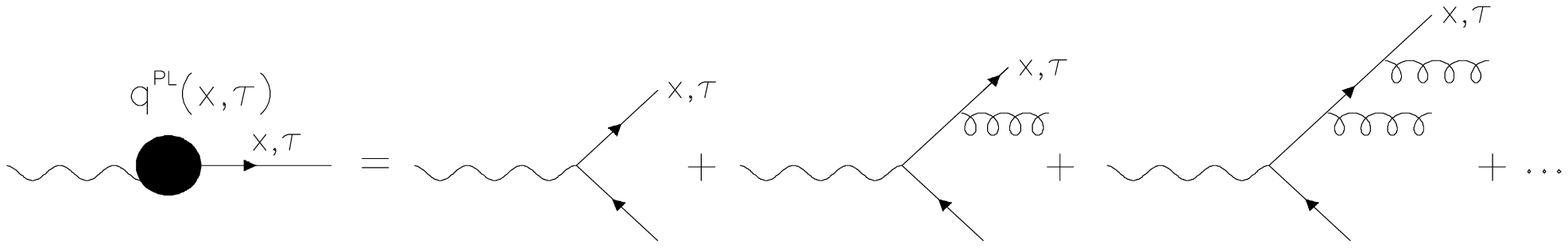,width=\textwidth}}
\put(0,0){\epsfig{file=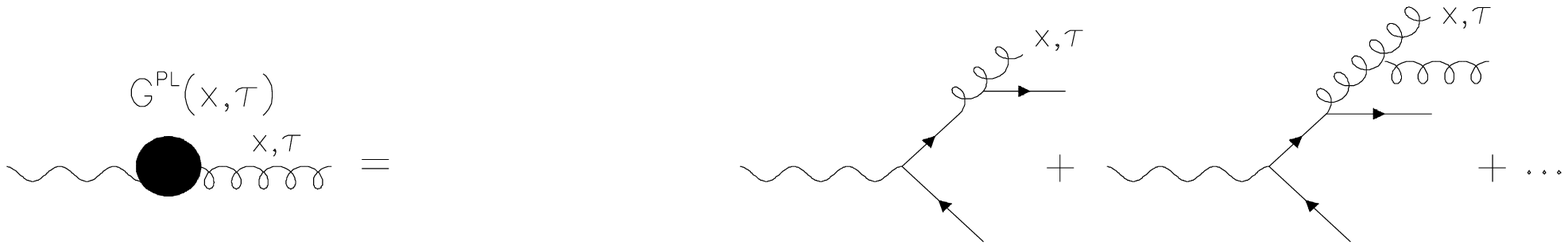,width=\textwidth}}
\end{picture}
\caption{Diagrams defining pointlike parts of nonsinglet
quark and gluon distribution functions of the photon in LL
approximation.}
\label{figpl}
\end{figure}
To see the most important feature of the pointlike part of quark
 distribution functions that will be crucial for the
following analysis, we now consider in detail the case of
nonsinglet quark distribution function $q_{\mathrm {NS}}^i(x,M)$ at
the LO (which is sufficient for our purposes) and, moreover, drop
the superscript $i$. A subset of all
pointlike solutions of (\ref{NSevolution}) may be characterized
by the value of the
initial scale $M_0$ at which they vanish. At the LO
and in terms of moments such solutions are given explicitly as
\begin{equation}
q_{\mathrm {NS}}^{\mathrm {PL}}(n,M_0,M)=\frac{4\pi}{\alpha_s(M)}
\left[1-\left(\frac{\alpha_s(M)}{\alpha_s(M_0)}\right)^
{-2P^{(0)}_{qq}(n)/\beta_0}\right]a_{\mathrm {NS}}(n),
\label{generalpointlike}
\end{equation}
where
\begin{equation}
a_{\mathrm {NS}}(n)\equiv \frac{\alpha}{2\pi\beta_0}
\frac{k_{\mathrm {NS}}^{(0)}(n)}{1-2P^{(0)}_{qq}(n)/\beta_0}.
\label{ans}
\end{equation}
and result from resummation of infinite series of diagrams
in the upper part of Fig. \ref{figpl}:
\begin{figure}\centering
\epsfig{file=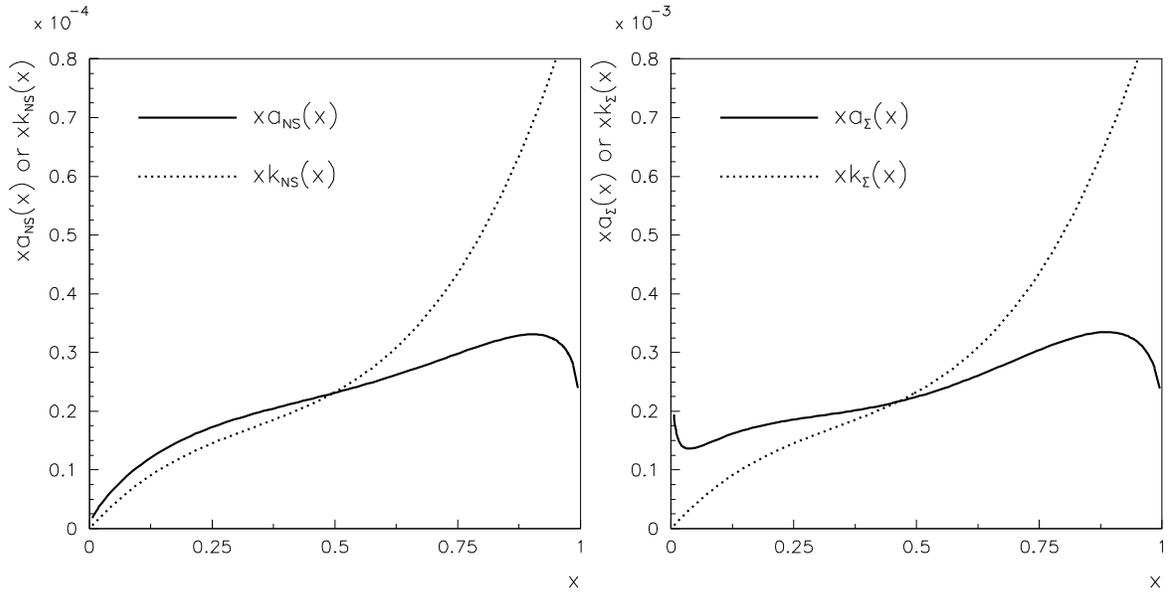,height=8cm}
\caption{Comparison, left in the nonsinglet and right in the singlet
channels,
of the functions $xk(x)$ given by the splitting term
$(\alpha/2\pi)3e_q^2(x^2+(1-x)^2)$, with the functions $xa(x)$
corresponding to asymptotic pointlike solutions (\ref{asymptotic}).}
\label{anskns}
\end{figure}
\begin{displaymath}
q^{\mathrm {PL}}_{\mathrm {NS}}(x,M_0,M) \equiv
\frac{\alpha}{2\pi}k_{\mathrm {NS}}^{(0)}(x)
\int^{M^2}_{M_0^2}\frac{{\mathrm d}\tau}{\tau}+
\int^{1}_{x}\frac{{\mathrm d}y}{y}P^{(0)}_{qq}
\left(\frac{x}{y}\right)
\int^{M^2}_{M_0^2}
\frac{{\mathrm d}\tau_1}{\tau_1}\alpha_s(\tau_1)
\frac{\alpha}{2\pi} k_{\mathrm {NS}}^{(0)}(y)\int^{\tau_1}_{M_0^2}
\frac{{\mathrm d}\tau_2}{\tau_2}+
\end{displaymath}
\begin{equation}
\int^{1}_{x}\frac{{\mathrm d}y}{y}P^{(0)}_{qq}
\left(\frac{x}{y}\right)
\int^{1}_{y}\frac{{\mathrm d}w}{w}P^{(0)}_{qq}
\left(\frac{y}{w}\right)
\int^{M^2}_{M_0^2}
\frac{{\mathrm d}\tau_1}{\tau_1}\alpha_s(\tau_1)
\int^{\tau_1}_{M_0^2}
\frac{{\mathrm d}\tau_2}{\tau_2}\alpha_s(\tau_2)
\frac{\alpha}{2\pi} k_{\mathrm {NS}}^{(0)}(w)\int^{\tau_2}_{M_0^2}
\frac{{\mathrm d}\tau_3}{\tau_3}
+\cdots.
\label{resummation}
\end{equation}
which, as illustrated in Fig. 2, softens substantially the
$x-$dependence of $a_{\mathrm {NS}}(x)$ with respect to the first
term in (\ref{resummation}), proportional to $k_{\mathrm {NS}}(x)$.
This construction is similar to that of hadrons with the first term
in (\ref{resummation}), $(\alpha/2\pi) k^{(0)}_{\mathrm
{NS}}(x)\ln(M^2/M_0^2)$, playing the role of ``bare'' quark
distribution function. In the case of the pointlike part
(\ref{resummation}) this bare distribution does, however, depend on
the scale $M$, and its derivative with respect to $\ln M^2$
generates the inhomogeneous term in (\ref{NSevolution}). Formally
$q^{\mathrm {PL}}_{\mathrm {NS}}(x,M_0,M)$ as given in
(\ref{generalpointlike},\ref{resummation}) can be considered also
for $M<M_0$, but is negative there. For $M/M_0\rightarrow
\infty$ the second term in brackets of (\ref{generalpointlike})
vanishes and therefore all the pointlike solutions share the same
large $M$ behaviour
\begin{equation}
q^{\mathrm {PL}}_{\mathrm {NS}}(x,M_0,M)
\rightarrow \frac{4\pi}{\alpha_s(M)}a_{\mathrm {NS}}(x)\equiv
q^{\mathrm {AP}}_{\mathrm {NS}}(x,M)\propto \ln \frac{M^2}{\Lambda^2},
\label{asymptotic}
\end{equation}
defining the so called {\em asymptotic pointlike} solution
$q^{\mathrm {AP}}_{\mathrm {NS}}(x,M)$ \cite{witten,BB}. Note that this
asymptotic pointlike solution is a very special case of the general
pointlike one (\ref{generalpointlike}), because for this solution
 the lower integration limit $M_0$ in (\ref{resummation}) has
been identified with $\Lambda$, i.e. $M_0=\Lambda$! The fact that
for the asymptotic pointlike solution (\ref{asymptotic})
$\alpha_s(M)$ appears in the denominator of (\ref{asymptotic}) has
been the source of claims (see, for instance, \cite{Vogt}) that
$q(x,M)={\cal O}(1/\alpha_s)$. This claim is wrong for two reasons.
First, it is obviously invalid for those of the currently used
parameterizations that distinguish the pointlike and hadronic
components of photonic PDF. For instance, the widely used
Schuler--Sj\"{o}strand sets SaS1 and SaS2, which take $M_0=0.6$ GeV
and $M_0=2$ GeV, respectively are not
asymptotic pointlike solutions and therefore manifestly do not
behave as $\alpha/\alpha_s$. However, as argued in \cite{jch1} this
claim is misleading even for the asymptotic pointlike solution. To
see the point consider the limit $\Lambda\rightarrow 0$ for fixed
$M_0$. We easily see that
\begin{equation}
 q^{\mathrm {PL}}_{\mathrm {NS}}(n,M,M_0)\rightarrow
\frac{\alpha}{2\pi} k_{\mathrm {NS}}(n,M)\ln\frac{M^2}{M_0^2},
\label{limits}
\end{equation}
i.e. (\ref{generalpointlike}) reduces to moments of the first term
in the expansion (\ref{resummation}), corresponding to pure QED
splitting $\gamma\rightarrow q\overline{q}$. The fact that the
asymptotic pointlike solution (\ref{asymptotic}), for which
$M_0=\Lambda$, diverges when $\Lambda\rightarrow 0$ is then a
direct consequence of the fact that for this (and only this)
pointlike solution the decrease of the coupling
$\alpha_s(M/\Lambda)$ as $\Lambda\rightarrow 0$ is overridden by the
simultaneous extension of the integration region as $M_0\rightarrow0$!
In other words, if QCD is switched off by sending
$\Lambda\rightarrow 0$ {\em without} simultaneously extending the
integration region, i.e. for {\em fixed $M_0$}, there is no trace
of QCD left and we get back the simple QED formula.

In summary, the pointlike part $q^{\mathrm {PL}}_{\mathrm {NS}}(x,M)$
of the quark distribution function of the photon is of the order
$\alpha$ and not of the order $\alpha/\alpha_s$, as often claimed.
One can use the latter only as a shorthand for the specification of
large $M$ behaviour as expressed in (\ref{asymptotic}). In fact,
this is what one finds in the original papers \cite{witten,BB}
\footnote{See, for instance, eq. (3.20) of \cite{BB}}
which do not contain any explicit claim that
$q(x,M)={\cal O}(1/\alpha_s)$.
To take this behaviour literally leads, as shown in the next
Section, to incorrect conclusions.

\section{Factorization scheme analysis of $F_2^{\gamma}(x,Q^2)$}
First a general comment. All the splitting and coefficient
functions discussed below can be calculated using one of two
different approaches
\begin{itemize}
\item Ultraviolet renormalization of composite operators within OPE.
\item Infrared renormalization technique anchored in the framework of
parton model and based on analysis of Feynman diagrams \cite{FP}.
\end{itemize}
For the discussion of the transition between the properties of real
and virtual photon, discussed in the next Section, the second
approach has clear advantage in its simplicity and physical
transparency and we shall therefore adopt it also throughout the
paper. Moreover we shall work with nonzero quark masses $m_q$,
which provide natural regulators of parallel singularity associated
with the primary $\gamma\rightarrow q\overline{q}$ splitting. A
simple and straightforward analysis of this splitting (see next
Section) yields
\begin{equation}
C_{\gamma}^{(0)}(x,Q/M)=3
\left[(x^2+(1-x)^2)\left(\ln\frac{M^2}{Q^2}+
\ln\frac{1-x}{x}\right)+
8x(1-x)-1\right]
\label{c0gamma}
\end{equation}
Note that when evaluated via ultraviolet renormalization of
composite operators using dimensional regularization
$C_{\gamma}^{(0)}$ contains also a term proportional to $\ln
4\pi-\gamma_E$. However, this term has nothing to do with QCD and
is related exclusively to the renormalization of electromagnetic
couplant $\alpha$\footnote{See also discussion following eq.
(\ref{redefinition}) below.}.

The reason why the existing NLO analyses of $F_2^{\gamma}(x,Q^2)$,
like \cite{Vogt,Aurenche,GRV1,GS,GRS,GRS2}, are incomplete is that
they all take literally the relation $q\propto 1/\alpha_s$. This, in
turn, leads to several incorrect conclusions:
\begin{itemize}
\item The term $C_{\gamma}^q$ in (\ref{f2gamma}), which starts
at the order ${\cal O}(\alpha)$, is claimed to be of the NLO in
$\alpha_s$ with respect to $q(x,M)$, while in fact it is of the
same order as $q(x,M)$.
\item The term in (\ref{cg}) proportional to
$\alpha C^{(0)}_{\gamma}$ is retained (though misleadingly
assigned to NLO), while the term proportional to $\alpha\alpha_s
C^{(1)}_{\gamma}$ is discarded, while in complete NLO analysis it
must be retained as well.
\item Similarly in the case of splitting functions
(\ref{splitquark}--\ref{splitpij}): while in the homogeneous
splitting functions terms proportional to $\alpha_s^2 P^{(1)}_{ij}$
are retained, the inhomogeneous splitting functions of the same
order, $\alpha_s^2 k^{(2)}_q,\alpha_s^2 k^{(2)}_G$ are discarded,
while they should be kept as well.
\end{itemize}
As a result, all existing analyses write $F_2^{\gamma}$ at
the LO in the form
\begin{equation}
F_2^{\gamma,{\mathrm {LO}}}(x,Q^2)=\sum_{q}2x e_q^2 q(x,M),
\label{wrongLO}
\end{equation}
where $q(x,M)$ satisfy LO evolution equations, which take into account
only $k_q^{(0}$ and $P_{ij}^{(0)}$ splitting functions and at the NLO
use the expression
\begin{eqnarray}
\lefteqn{F_2^{\gamma,{\mathrm {NLO}}}(x,Q^2)= \sum_{q}2x e_q^2 q(x,M)+}
\label{wrongNLO}
\\ & &
\sum_{q}2xe_q^2\left(\frac{\alpha_s(M)}{2\pi}\left[q(M)\otimes
C_q^{(1)}(Q/M)+ G(M)\otimes
C_G^{(1)}(Q/M)\right]+
\frac{\alpha}{2\pi}e_q^2C^{(0)}_{\gamma}(x,Q/M)\right),
\nonumber
\end{eqnarray}
where the evolution equations for quark and gluon distribution
functions include terms up to $k^{(1)}_q,k^{(1)}_G$ and
$P^{(1)}_{ij}$. However, taking into account that
$q(x,M)=O(\alpha)$, correct LO and NLO
expressions for $F_2^{\gamma}$  read instead as follows
\begin{equation}
F_2^{\gamma,{\mathrm {LO}}}(x,Q^2)=\sum_{q}2x e_q^2
\left(q(x,M)+\frac{\alpha}{2\pi}e_q^2C^{(0)}_{\gamma}(x,Q/M)\right),
\label{rightLO}
\end{equation}
\begin{eqnarray}
\lefteqn{F_2^{\gamma,{\mathrm {NLO}}}(x,Q^2)= \sum_{q}2x e_q^2
\left(q(x,M)+\frac{\alpha}{2\pi}e_q^2C^{(0)}_{\gamma}(x,Q/M)\right)+}
\label{rightNLO}
\\ & &
\sum_{q}2xe_q^2\left(\frac{\alpha_s(M)}{2\pi}\left[q(M)\otimes
C_q^{(1)}(Q/M)+ G(M)\otimes
C_G^{(1)}(Q/M)\right]+
\frac{\alpha}{2\pi}\frac{\alpha_s(M)}{2\pi}
e_q^2C^{(1)}_{\gamma}(Q/M)\right), \nonumber
\end{eqnarray}
where PDF in (\ref{rightNLO}) satisfy
evolution equations with splitting functions
(\ref{splitquark}--\ref{splitpij}) up to the
order $\alpha_s^2(M)$, i.e. including terms proportional to
$k^{(2)}_q,k^{(2)}_G$. The
difference of (\ref{rightLO}--\ref{rightNLO}) with respect to
(\ref{wrongLO}--\ref{wrongNLO}) is threefold:
\begin{itemize}
\item The term proportional to $C^{(0)}_{\gamma}$ is part of
the LO expression.
\item The NLO expression in (\ref{rightNLO}) contains, beside the
usual partonic convolutions $q\otimes C^{(1)}_q$ and
$G\otimes C^{(1)}_G$
also the NLO direct term proportional to $C^{(1)}_{\gamma}$.
This additional term is crucial for
the consistency of the NLO approximation. As shown below, it also
cancels part of the factorization scale and scheme dependence of
the NLO hadronic terms and thus contributes to theoretical
stability of NLO calculations.
\item In the NLO evolution equations inhomogeneous splitting
functions $k_i^{(2)},i=q,G$ of the order $\alpha_s^2$ are retained.
\end{itemize}
We note that in the case of the proton the expressions for
$F_2^{\mathrm p}(x,Q^2)$ at LO and NLO differ from those in
(\ref{rightLO}-\ref{rightNLO}) merely by the absence of the terms
proportional to $C^{(0)}_{\gamma}$ and $C^{(1)}_{\gamma}$. For
$F_2^{\mathrm p}(x,Q^2)$ factorization scale and scheme dependence
of parton distribution functions is cancelled by explicit
factorization scale and scheme dependence of the coefficient
functions $C^{(i)}_q,C^{(i)}_G,i\ge 1$. Concretely, at the NLO the
scale and scheme dependence of quark distribution functions in the
first (i.e. LO) term on the r.h.s. of (\ref{rightNLO}) is cancelled
to order $\alpha_s$ by the corresponding dependence of NLO
coefficient coefficients $C^{(1)}_q,C^{(1)}_G$. This reflects the
fact that logarithmic derivatives of PDF of the proton with respect
to $\ln M$ and FS start at the order $\alpha_s$ and also implies
that at the LO the relation between $F_2^{\mathrm p}(x,Q^2)$ and
quark distribution functions is {\em unique}. Moreover, a precise
physical interpretation of the factorization scale $M$ is of little
relevance because any redefinition of $M$ can be compensated by
appropriate change of the NLO splitting functions $P^{(1)}_{kl}$.

For photon the inhomogeneous term in evolution equations for quark
distribution functions modifies this cancellation mechanism because
part of the factorization scale dependence driven by this
inhomogeneous term is compensated already at the LO by the term in
$C^{(0)}_{\gamma}$ proportional to $k^{(0)}_q\ln(Q^2/M^2)$! The
presence of such a term in $C^{(0)}_{\gamma}$, as well as the fact
that $C^{(1)}_{\gamma}$ contains the term proportional to
$k^{(1)}_q$, are general consequences of the RG invariance of
$F_2^{\gamma}(x,Q^2)$. For photon the interpretation of
factorization scale $M$ therefore does matter, as it determines LO
coefficient $C^{(0)}_{\gamma}$. Note that $M^2$ in the
(\ref{c0gamma}) must be interpreted as maximal transverse momentum
squared of the virtual quark included in the resummation
(\ref{resummation}).

To see explicitly how the cancelation of factorization scheme
dependence works at the NLO and why it requires the inclusion of
the term $C^{(1)}_{\gamma}$, let us follow the standard procedure
\cite{Vogt} of redefining quark distribution functions of the
photon, but include terms up to the order $\alpha\alpha_s$:
\begin{equation}
\tilde{q}(x,M)\equiv q(x,M)+\frac{\alpha}{2\pi}e_q^2D_{\gamma}(x)+
\frac{\alpha}{2\pi}\frac{\alpha_s(M)}{2\pi}e_q^2 E_{\gamma}(x).
\label{redefinition}
\end{equation}
It is straightforward to show that provided $q(x,M)$ satisfies the
evolution equation with inhomogeneous splitting functions
$k_q^{(0)}, k_q^{(1)}$ and $k^{(2)}_q$, the tilded distribution
function $\tilde{q}(x,M)$ satisfies the same evolution equation,
but with tilded inhomogeneous splitting and coefficient functions
given as:
\begin{eqnarray}
\tilde{k}_q^{(0)} & = & k_q^{(0)}, \label{tildek0}\\
\tilde{k}_q^{(1)} & = & k_q^{(1)}-e_q^2P^{(0)}_{qq}
\otimes D_{\gamma}, \label{tildek1}\\
\tilde{k}_q^{(2)} & = & k_q^{(2)}-
e_q^2\left(P^{(1)}_{qq}\otimes D_{\gamma}+P^{(1)}_{qq}\otimes
E_{\gamma}+\beta_0 E_{\gamma}/2\right).
\label{tildek2} \\
\tilde{C}^{(0)}_{\gamma} & = & C^{(0)}_{\gamma}-D_{\gamma},
\label{tildec0}\\
\tilde{C}^{(1)}_{\gamma} & = &
C^{(1)}_{\gamma}- C^{(1)}_q\otimes D_{\gamma}-E_{\gamma}.
\label{tildec1}
\end{eqnarray}
On the other hand, NLO homogeneous splitting functions
$P^{(1)}_{ij}$ are unchanged by the redefinition
(\ref{redefinition}). The relations (\ref{tildek1}--\ref{tildek2})
imply that instead of the functions $D_{\gamma}(x)$ and
$E_{\gamma}(x)$ this redefinition can equally well be parameterized
by the inhomogeneous splitting functions $k^{(1)}_q,k^{(2)}_q$.

Despite the fact that $D_{\gamma}$ is related by (\ref{tildek1}) to
the splitting function $k^{(1)}_q$ standing in (\ref{splitquark})
by $\alpha_s$, only the term proportional to
$\alpha\alpha_sE_{\gamma}$ is related to genuine QCD effects. This
is indicated already by the observation that solutions of the
equation
\begin{equation}
\frac{{\mathrm d}q(x,M)}{{\mathrm d}\ln M^2}=
\frac{\alpha}{2\pi}k^{(0)}_q(x)
\label{qedevolution}
\end{equation}
to which the evolution equation (\ref{NSevolution}) reduces in the
limit $\alpha_s\rightarrow 0$ are determined up to an arbitrary
function of $x$, i.e. just like the term in (\ref{redefinition})
proportional to $D_{\gamma}(x)$. The fact that this term in
(\ref{redefinition}) has nothing to do with QCD follows also from
the fact that it can be fully included in the redefinition of the
lower integration bound $M_0^2$ in (\ref{resummation}) by setting
$M_0^2(x)=M_0^2\exp(-D_{\gamma}(x)/(x^2+(1-x)^2))$. With such a
choice of the lower integration bound the result of the resummation
in (\ref{resummation}) does not vanish at any fixed initial scale,
but still solves the same inhomogeneous evolution equation and
therefore represents legitimate pointlike solution.

It is also illustrative to see how the redefinition
(\ref{redefinition}) affects separately the pointlike and hadronic
parts of quark distribution functions. For the term containing
$D_{\gamma}$ the answer can be taken from the analysis in
\cite{Aurenche}, where explicit expressions for moments
$F_2^{\gamma}(n,M^2)$ as functions of $k^{(1)}_q(n)$ are given. One
finds that the variation of $k^{(1)}_q$ by $\delta k^{(1)}_q\equiv
\tilde{k}_q^{(1)}-k_q^{(1)}=
-e_q^2P^{(0)}_{qq}\otimes D_{\gamma}$ modifies
the pointlike part $q^{\mathrm {PL}}(n,M)$ at all scales $M$ by the
term ($d(n)\equiv 2P^{(0)}_{qq}/\beta_0$)
\begin{equation}
\delta q^{\mathrm {PL}}(n,M)=
-\frac{\alpha}{2\pi}\left[1-\left(\frac{\alpha_s(M)}{\alpha_s(M_0)}
\right)^{-d(n)}\right]
\frac{\delta k^{(1)}_q}{P^{(0)}_{qq}(n)}=
\frac{\alpha}{2\pi}\left[1-\left(\frac{\alpha_s(M)}{\alpha_s(M_0)}
\right)^{-d(n)}\right]e_q^2D_{\gamma}(n),
\label{deltaPL}
\end{equation}
while in the hadronic part only the initial condition at $M_0$ is
changed
\begin{equation}
\delta q^{\mathrm {HAD}}(n,M_0)=-\frac{\alpha}{2\pi}
\frac{\delta k^{(1)}_q}{P^{(0)}_{qq}(n)}=\frac{\alpha}{2\pi}e_q^2
D_{\gamma}(n).
\label{deltaHAD}
\end{equation}
The NLO splitting functions remain, however, unchanged. Summing
(\ref{deltaPL}) with (\ref{deltaHAD}) multiplied by the evolution
factor $(\alpha_s(M)/\alpha_s(M_0))^{-d(n)}$ yields the second term
on the r.h.s. of (\ref{redefinition}).

\section{Structure of the virtual photon}

\subsection{Equivalent photon approximation}
All present knowledge of the structure of the photon comes from
experiments at the ep and e$^+$e$^-$ colliders, where incoming
leptons act as sources of transverse and longitudinal virtual
photons
\footnote{In a typical ``photoproduction'' experiment at HERA the
average photon virtuality $\langle P^2\rangle\doteq
10^{-2}-10^{-3}$ GeV$^2$.}. To order $\alpha$ their respective
unintegrated fluxes are given as
\begin{eqnarray}
f^{\gamma}_{T}(y,P^2) & = & \frac{\alpha}{2\pi}
\left(\frac{1+(1-y)^2)}{y}\frac{1}{P^2}-\frac{2m_{\mathrm e}
^2 y}{P^4}\right),
\label{fluxT} \\
f^{\gamma}_{L}(y,P^2) & = & \frac{\alpha}{2\pi}
\frac{2(1-y)}{y}\frac{1}{P^2}.
\label{fluxL}
\end{eqnarray}
The transverse and longitudinal fluxes thus coincide for $y=0$,
while at $y=1$, $f_{L}^{\gamma}$ vanishes. Note that while for
$P^2\gg m_{\mathrm e}^2$ the second term in (\ref{fluxT}) is
negligible with respect to $1/P^2$, for $P^2$ close to
$P^2_{\mathrm min}=m^2 y^2/(1-y)$ their ratio is finite and
approaches $2(1-y)/(1+(1-y)^2)$.

\subsection{From virtual to real photons}
The choice of a way mass singularities resulting from the primary
$\gamma^*\rightarrow q\overline{q}$ splitting are regularized is
crucial for smooth and physically transparent transition between
the properties of virtual and real photons.  As far as perturbative
calculations of coefficient functions (hard scattering
cross--sections in general) are concerned, there is no principal
difference in this respect between QED and QCD. For PDF of the
photon the situation is, however, more complicated. The point is
that in QCD the transition from virtual to real photon  is expected
to be determined by nonperturbative parameters related to colour
confinement, rather than the quark masses, which govern such
transition in pure QED. Nevertheless, it is illustrative to see how
this transition is realized in QED, where the ``lepton distribution
functions'' of the photon, are explicitly calculable.

In QED masses of charged fermions, in particular electron, play a
fundamental role. For real photon parallel singularity associated
with the purely QED splitting $\gamma\rightarrow q \overline{q}$ is
naturally regulated by quark and lepton masses. For virtual photon
also its virtuality $P^2$ shields off this singularity but quark
and lepton masses are indispensable for proper limiting behaviour
of lepton (quark) distribution functions of the virtual photon as
$P^2\rightarrow 0$. The effects of nonzero photon virtuality $P^2$
 are threefold:
\begin{itemize}
\item longitudinally polarized photons must be taken into account,
\item unintegrated PDF of transversally polarized photons
contain terms proportional to $P^2$,
\item parton level cross--sections obtain contributions from terms
proportional to $P^2$.
\end{itemize}

\subsection{Parton model interpretation of constant terms
in $C^{(0)}_{\gamma}$}
\begin{figure}\centering
\epsfig{file=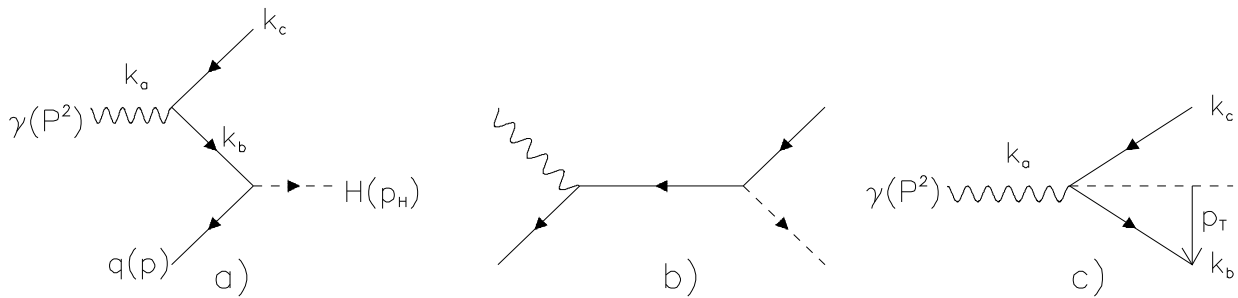,width=\textwidth}
\caption{Kinematics of the basic
$\gamma^*(P^2)\rightarrow q\overline{q}$ splitting in QED.}
\label{gammaffbar}
\end{figure}
The evaluation of quark distribution functions of the virtual
photon in pure QED serves as a useful guide to parton model
interpretation of the so called ``constant terms'' in LO
coefficient function $C^{(0)}_{\gamma}$ and leads to explicit
formulae for the virtuality dependence of quark distribution
functions of the photon.

Consider, for definiteness, the production of a heavy scalar
particle $H$ with mass $M_H$ in electron--proton collisions. The
relevant parton level hard scattering subprocess
\begin{equation}
\gamma(P^2,k_a)+q(p)\rightarrow {\mathrm e}(k_c)+H(p_H)
\label{hard}
\end{equation}
is described by the sum of amplitudes corresponding to diagrams in
Fig. 3a,b. In pure QED and to order $\alpha$ the probability of
finding inside the photon of virtuality $P^2$ a quark with mass
$m_q$, electric charge $e_q$, momentum fraction $x$ and virtuality
$\tau=m_q^2-k_b^2$ up to $M^2$ is given by the t--channel diagram
in Fig. 3a
\footnote{The s--channel diagram in Fig. 3b is, however,
crucial for preserving gauge invariance in the case of the
longitudinal photon where it cancels one of the terms originating
from Fig. 3a that does not vanish for $P^2\rightarrow 0$.}
\begin{equation}
q_{\mathrm {QED}}(x,m^2_q,P^2,M^2)=
\left(\frac{\alpha}{2\pi}3e_q^2\right)
\int_{\tau^{\mathrm {min}}}^{M^2}
\frac{W_t(x,m^2_q,P^2)}{\tau^2}{\mathrm d}\tau
\label{general}
\end{equation}
In the collinear kinematics, which is relevant for finding the
lower limit on $\tau$, the values of $m_q,~x,~\tau$ and $p_T$ are
related to initial photon virtuality $P^2$ as follows
\begin{equation}
\tau=xP^2+\frac{m_q^2}{1-x}+p_T^2,~~~
\Rightarrow ~~~\tau^{\mathrm {min}}=xP^2+\frac{m_q^2}{1-x}.
\label{kinematics}
\end{equation}
In terms of $M_H$ and $s\equiv (k_a+p)^2$, $x=M_H^2/s$.
$W(x,m^2,P^2,M^2)$ can in general be written as
\begin{eqnarray}
 W(x,m^2_q,P^2,M^2)& =
&f(x)\frac{p_T^2}{1-x}+g(x)m^2_q+
h(x)P^2+c(x)\frac{\tau^2}{s}+\cdots
\nonumber \\
& = & f(x)\tau+\left(g(x)-\frac{f(x)}{1-x}\right)m^2_q+
\left(h(x)+xf(x)\right)P^2+c(x)\frac{\tau^2}{s}+\cdots
\label{W}
\end{eqnarray}
where the dots indicate term of the type $\tau^{k+1}/s^k,k\ge 1$.
There is no term proportional to $s$ as it would violate unitarity.
The functions $f(x),g(x)$ and $h(x)$ are unique functions that can
be determined from the analyses of the vertex $\gamma\rightarrow
q\overline{q}$ in collinear kinematics. On the other hand, $c(x)$
is a process dependent. The terms in (\ref{W}) proportional to
$f(x),g(x)$ and $h(x)$ are dominated by small quark virtualities
and have therefore clear parton model interpretation: so long as
$\tau\ll M_H^2$ eq. (\ref{general}) describes the flux of quarks
that are almost collinear with the incoming photon and ``live''
long with respect to the production time of the heavy particle $H$
in the lower vertex in Fig. 3a. Substituting (\ref{W}) into
(\ref{general}) and performing the integration gives, in units of
$3e_q^2\alpha/2\pi$,
\begin{equation}
q_{\mathrm {QED}}(x,m^2_q,P^2,M^2)=
f(x)\ln\left(\frac{M^2}{\tau^{\mathrm {min}}}\right)+
\left[-f(x)+\frac{g(x)m^2_q+h(x)P^2}{\tau^{\mathrm {min}}}
\right]
\left(1-\frac{\tau^{\mathrm {min}}}{M^2}\right)+c(x)\frac{M^2}{s}.
\label{fullresult}
\end{equation}
In practical applications the factorization scale M is identified
with some kinematical variable characterising hardness of the
interaction, like $Q^2$ in DIS or $M_H$ in our process
(\ref{hard}). As a result $M^2/s$ becomes a function of $x$ and we
can write $c(x)M^2/s=\kappa(x)$. For $\tau^{\mathrm
min}=xP^2+m^2_q/(1-x)\ll
 M^2$ (\ref{fullresult}) simplifies to
\begin{equation}
q_{\mathrm {QED}}(x,m^2_q,P^2,M^2)  =  f(x)\ln\left(
\frac{M^2}{xP^2+m^2_q/(1-x)}\right)-f(x)+
\frac{g(x)m^2+h(x)P^2}{xP^2+m^2_q/(1-x)}+\kappa(x).
\label{finalresult}
\end{equation}
This basic result, which takes into account nonzero quark mass
$m_q$ as well as initial photon virtuality $P^2$, reduces for
$x(1-x)P^2\gg m^2$ to
\begin{equation}
q_{\mathrm {QED}}(x,0,P^2,M^2)=f(x)\ln\left(\frac{M^2}{xP^2}\right)
-f(x)+\frac{h(x)}{x}+\kappa(x),
\label{virtualphoton}
\end{equation}
whereas for $P^2/m_q^2\rightarrow 0$ it approaches
\begin{equation}
q_{\mathrm {QED}}
(x,m^2_q,0,M^2)=f(x)\ln\left(\frac{M^2(1-x)}{m^2_q}\right)
-f(x)+g(x)(1-x)+\kappa(x),
\label{realphoton}
\end{equation}
describing distribution of quarks inside the real photon. The fact
that (\ref{realphoton}) requires nonzero quark mass reflects the
fact that in QED masses of charged fermions play vital role and
cannot be zero.

As in the case of the photon fluxes (\ref{fluxT}-\ref{fluxL})
 the leading logarithmic term, dominant for large $M^2$,
as well as the ``constant'' terms proportional to $f(x),g(x)$ and
$h(x)$ come entirely from the integration region close to
$\tau^{\mathrm {min}}$ and are therefore unique. At
$\tau=\tau^{\mathrm {min}}$ both types of the singular terms in
(\ref{general}), i.e. $1/\tau$ and $m_q^2/\tau^2$ or $P^2/\tau^2$,
are of the same order but the faster fall--off of the $1/\tau^2$
terms implies that for large $M^2$ the integral in (\ref{general})
is dominated by the weaker singularity $1/\tau$. In other words,
while the logarithmic term is dominant at large $M^2$, the constant
terms resulting from nonzero $m^2$ and/or $P^2$ come from the
kinematical configurations which are even more collinear than those
giving the logarithmic term. On the other hand, not all constant
terms are of this origin, as exemplified by the term proportional
to $\kappa(x)$ which comes from the integration over the whole
phase space and is therefore process dependent. These terms have no
parton model interpretation and belong therefore naturally to the
coefficient function $C^{(0)}_{\gamma}$.

The analysis of the vertex $\gamma(P^2)\rightarrow q\overline{q}$
in collinear kinematics or the explicit evaluation of the diagrams
in Fig. \ref{gammaffbar} yields the following results
\begin{equation}
\begin{array}{lll}
 f_T(x)=x^2+(1-x)^2, & g_T(x)=\frac{1}{1-x}, & h_T(x)=0, \\
 f_L(x)=0, & g_L(x)=0, & h_L(x)=4x^2(1-x).
\label{fghTL}
\end{array}
\end{equation}
The vanishing of $f_L(x)$ and $g_L(x)$ is a consequence of gauge
invariance, which guarantees that $\gamma_L$ decouples in the limit
$P^2\rightarrow 0$. The fact that $h_T(x)=0$ is due to helicity
conservation, which permits violation by quark mass terms only. The
expressions (\ref{fullresult}-\ref{finalresult}) exhibit explicitly
the smooth transition between the quark distribution functions of
the virtual and real photon in pure QED, governed by the ratio
$P^2/m^2_q$.

For virtual photon eq. (\ref{f2gamma}) holds separately for both
transverse and longitudinal polarizations. For $x(1-x)P^2\gg m^2$
the finite terms $C_{\gamma}^T,C_{\gamma}^L$ are given as
\cite{russians}
\begin{eqnarray}
C_{\gamma,T}^{(0)}(x,P^2,Q/M)&
= & 3
\left[(x^2+(1-x)^2)\left(\ln\frac{M^2}{Q^2}+\ln\frac{1}{x^2}\right)+
8x(1-x)-2\right],
\label{cgammaTvirtual} \\
C_{\gamma,L}^{(0)}(x,P^2,Q/M)& = & 4x(1-x),
\label{cgammaLvirtual}
\end{eqnarray}
while for $P^2=0$
\begin{eqnarray}
C_{\gamma,T}^{(0)}(x,0,Q/M)& = &
3\left[(x^2+(1-x)^2)\left(\ln\frac{M^2}{Q^2}+
\ln\frac{1-x}{x}\right)+
8x(1-x)-1\right],
\label{cgammaTreal} \\
C_{\gamma,L}^{(0)}(x,0,Q/M)& = & 0
\label{cgammaLreal}
\end{eqnarray}
In combination with (\ref{finalresult}) and (\ref{fghTL}) these
results imply $\kappa(x)=-1+6x(1-x)$. The origins of the
nonlogarithmic parts of $C^{(0)}_{\gamma}$ in
(\ref{cgammaTvirtual}-\ref{cgammaTreal}) can then be identified as
follows:
\begin{description}
\item {\bf Real photon:}
\begin{eqnarray}
-1+8x(1-x) & = & \underbrace{-2+8x(1-x)}_{\mathrm {for~massless~quark}}+
\underbrace{1}_{g^T(x)(1-x)} \nonumber \\
 & = & \underbrace{-1+6x(1-x)}_{\mathrm {nonuniversal~part}}-
\underbrace{(x^2+(1-x)^2)}_{f^T(x)}+\underbrace{1}_
{g^T(x)(1-x)}~~~~~~~~~
\label{intreal}
\end{eqnarray}
\item{\bf Virtual photon:}
\begin{equation}
-2+8x(1-x)  = \underbrace{-1+6x(1-x)}_{\mathrm {nonuniversal~part}}
-\underbrace{(x^2+(1-x)^2)}_{f^T(x)}
\label{intvirtual}
\end{equation}
\end{description}
As the virtuality $P^2$ (or more precisely $\tau^{\mathrm min}$)
increases toward the factorization scale $M^2$, the expression
(\ref{fullresult}) for the quark distribution functions of the
virtual photon approaches zero. We emphasize that this holds not
only for the logarithmic term but for all terms that have parton
model interpretation.

\subsection{The real world}
In realistic QCD nonperturbative effects, in particular those
connected with the confinement, are expected to determine the
structure of the real photon as well as its virtuality dependence.
For instance, within the Schuler--Sj\"{o}strand set of
parameterizations \cite{sas1} the role of such parameter is played
by vector meson masses for the hadronic component and by the
initial $M_0$ for the pointlike one. Nevertheless, the analysis of
the previous subsections is still relevant for the coefficient
function $C_{\gamma}^{(0)}$ as well as the discussion of virtuality
dependence of the pointlike part of quark distribution functions.

\section{Phenomenological implications}
The results of previous Sections have several implications for
phenomenological analyses of hard scattering processes with photons
in the initial state. Some of them are discussed below.

\subsection{LO and NLO analysis of $F_2^{\gamma}(x,Q^2)$ for the
real photon}
\begin{figure}\centering
\epsfig{file=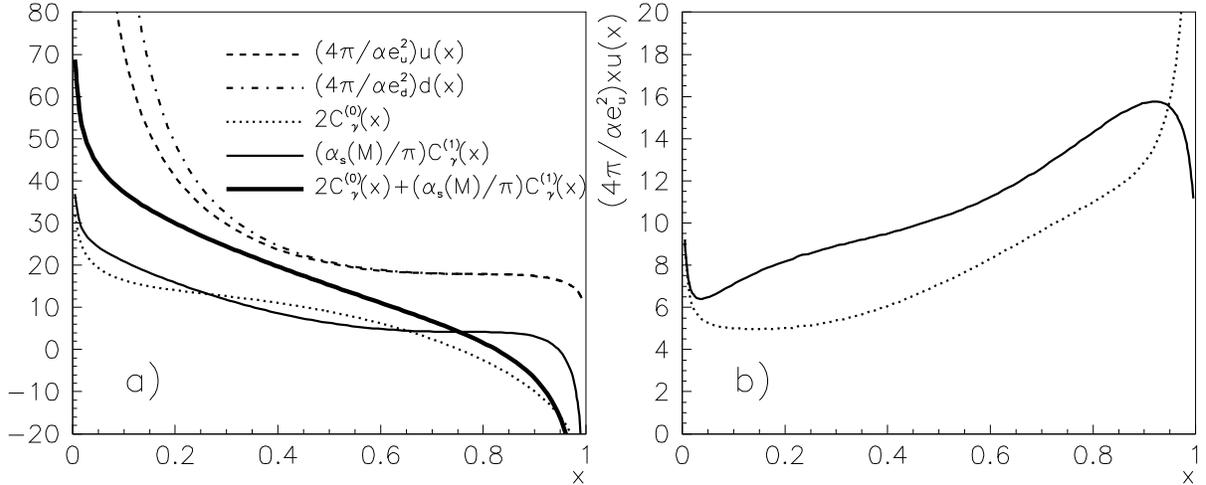,width=\textwidth}
\caption{Comparison of the contributions of
coefficient functions $C^{(0)}_{\gamma}(x)$ and
$C^{(1)}_{\gamma}(x)$ to $F_2^{\gamma}$ with those of quark
distribution functions $u(x)$ and $d(x)$ (a). In b) the SaS1M
parameterization of $xu(x)$ (solid line) is compared to the
quantity (\ref{c2effects}) (dotted curve).}
\label{c2fig}
\end{figure}
As argued in detail in Section 3, the existing analyses of
$F_2^{\gamma}(x,Q^2)$ are incomplete due to the fact that they do
not take into account
\begin{itemize}
\item the NLO photonic coefficient function $C_{\gamma}^{(1)}(x)$ in
expression (\ref{rightNLO}) for $F_2^{\gamma}$ an
\item the NLO inhomogeneous splitting functions $k_q^{(2)}$ and
$k_G^{(2)}$ in evolution equations
(\ref{Sigmaevolution}-\ref{NSevolution}).
\end{itemize}
There is no problem to remedy the first shortcoming, because the
coefficient function $C_{\gamma}^{(1)}(x)$ is actually known. As
argued in \cite{Willy1} it can be deduced from the ${\cal
O}(\alpha_s^2)$ gluonic coefficient function calculated in
\cite{Willy2}. For illustration of its numerical importance,
$(\alpha_s(M)/\pi)C_{\gamma}^{(1)}(x)$ in $\overline{\mathrm
MS}+$MS scheme is compared in Fig. \ref{c2fig}a to
$2C_{\gamma}^{(0)}(x)$ as well as to $(4\pi/\alpha e_q^2)u(x)$ and
$(4\pi/\alpha e_q^2)u(x)$ corresponding to SaS1M parameterization.
In this comparison we took
$\alpha_s(M)/2\pi=0.03$, which is approximately the value of
$\alpha_s$ at $M=10$ GeV. Note that the contribution of
$C^{(1)}_{\gamma}$ to $F_2^{\gamma}(x,Q^2)$ is comparable to that
of $C^{(0)}_{\gamma}$. As in most experiments $Q^2$ is well below
$100$ GeV$^2$, the relative importance of $C^{(1)}_{\gamma}$ is
even bigger. We also see
that for $x\sim 0.4$ sum of the contributions of coefficient
functions $C^{(0)}_{\gamma}$ and $C^{(1)}_{\gamma}$ to
$F_2^{\gamma}$ is almost the same as the contributions of quark
distribution functions themselves.

 As far as the NLO inhomogeneous splitting
functions $k_q^{(2)}(x)$ and $k_G^{(2)}(x)$ are concerned, the
situation is different. These functions are unknown as they cannot
be derived from the existing NLO calculations. To obtain them
requires complete NNLO calculation, similar to that in \cite{larin}
for moments of nucleon structure functions. No such calculation is,
however, in sight. In the absence of $k_q^{(2)}(x)$ and
$k_G^{(2)}(x)$ a complete NLO analysis of $F_2^{\gamma}$ is
impossible, but we may investigate what happens if at least the
effects of $C^{(1)}_{\gamma}$ are taken into account. The simplest
way to estimate these effects is to evaluate the difference
\begin{equation}
q'(x,M)\equiv q(x,M)-
\frac{\alpha}{2\pi}\frac{\alpha_s(M)}{2\pi}e_q^2C^{(1)}_{\gamma}(x),
\label{c2effects}
\end{equation}
defining quark distribution functions which when inserted into
(\ref{rightNLO}) including the NLO coefficient function
$C^{(1)}_{\gamma}$ return the same $F_2^{\gamma}$ as $q(x,M)$
inserted into (\ref{rightNLO}) without the $C^{(1)}_{\gamma}$ term.
For $u$--quark the effect of the redefinition (\ref{c2effects}) is
shown in Fig. \ref{c2fig}b, using SaS1M parameterization of $u(x,M)$
at $M=10$ GeV. The reduction of $u(x,M)$ due to the inclusion of
$C^{(1)}_{\gamma}$ is substantial for all $0.05\le x\le 0.9$.

In the situation when full NLO calculation of $F_2^{\gamma}$ is
impossible due to the lack of knowledge of $k_q^{(2)}$ and
$k_G^{(2)}$ and Fig. \ref{c2fig} indicates the importance of
$C^{(1)}_{\gamma}$ term, LO analyses of $F_2^{\gamma}$ and other
quantities involving initial photons is all we can consistently do.
In such analyses we can exploit the freedom in the choice of the
factorization scale and scheme. Contrary to the case of hadrons,
these choices influence the relation between PDF and $F_2^{\gamma}$
(and other physical quantities) already at the LO through
factorization scale and scheme dependence of $C_{\gamma}^{(0)}$. It
is therefore not true, as claimed for instance in \cite{marco},
that the SaS1M and SaS2M sets of parameterizations are
``theoretically inconsistent'' because they combine in LO
expression for $F_{2}^{\gamma}$ the ``NLO'' quantity
$C_{\gamma}^{(0)}$ with the LO quark distribution functions. The
$\overline{\mathrm {MS}}$ sets of SaS parameterizations are as
legitimate definitions of the LO quark distribution functions as
their DIS companions and there is no theoretical reason why, as
suggested in \cite{marco}, they should not be used in
phenomenological analyses.

\subsection{What is measured in DIS on virtual photons?}
In experiments \cite{PETRA,PEP,TRISTAN,LEP} at e$^+$e$^-$ colliders
structure of the photon was investigated via standard DIS on the
photon with small but nonzero virtuality $P^2$. The resulting data
were used in \cite{GRS,GRS2} to determine PDF of the virtual photon
in LO and NLO approximations. In these analyses $C^{(0)}_{\gamma}$
was taken in the form
\begin{equation}
C_{\gamma}^{(0)}(x,P^2,Q/M)=3
\left[(x^2+(1-x)^2)\left(\ln\frac{M^2}{Q^2}+\ln\frac{1}{x^2}\right)+
6x(1-x)-2\right],
\label{c0virtwrong}
\end{equation}
which, however, does not correspond to the structure function that
is really measured in e$^+$e$^-$ collisions, but to the following
combination
\begin{equation}
F_{2,\Sigma}^{\gamma}(x,P^2,Q^2)\equiv F_{2,T}^{\gamma}(x,P^2,Q^2)-
\frac{1}{2}F_{2,L}^{\gamma}(x,P^2,Q^2)
\label{fsigma}
\end{equation}
of structure functions corresponding to transverse and longitudinal
polarizations of the target photon. This combination results after
averaging over the target photon polarizations by means of
contraction with the tensor $-g_{\mu\nu}/2$. The term $-2+6x(1-x)$
follows also directly from considerations of the previous Section:
\begin{equation}
-2+6x(1-x) =  \underbrace{-2+8x(1-x)}_{{\mathrm {from}}~\gamma^T}-
\underbrace{2x(1-x)}_{{\mathrm {from}}~\gamma^L/2}
\label{intvirt}
\end{equation}
Because the fluxes (\ref{fluxT}--\ref{fluxL}) of transverse and
longitudinal photons are different functions of $y$, any complete
analysis of experimental data in terms of the structure functions
$F_{2,T}^{\gamma}(x,P^2,Q^2)$ and $F_{2,L}^{\gamma}(x,P^2,Q^2)$ at
fixed $x,P^2,Q^2$ requires combining data for different $y$.
This is in principle possible, but
experimentally difficult to accomplish. The situation is simpler at
small $y$, where $f_T^{\gamma}(y,P^2)\doteq
f_L^{\gamma}(y,P^2)=f^{\gamma}(y,P^2)$, and the data therefore
correspond to the convolution of $f^{\gamma}(y,P^2)$ with the sum
$F_{2,T}^{\gamma}+F_{2,L}^{\gamma}$. The nonlogarithmic term in
$C^{(0)}_{\gamma}$ appropriate to this combination is, however, not
$-2+6x(1-x)$, as in (\ref{c0virtwrong}) and \cite{GRS,GRS2} but
$-2+12x(1-x)$, the sum of nonlogarithmic terms corresponding to
transverse and longitudinal photons. Numerically the difference
between these two expressions is quite important.

\subsection{Longitudinal gluons inside hadrons?}
If there are longitudinal photons inside leptons, what about
longitudinal gluons inside hadrons? Although real longitudinal
gluons decouple, as do real longitudinal photons, gluons as well as
quarks inside hadrons are off--shell and therefore there in no
reason not to introduce distribution functions of transverse and
longitudinal gluons as well. The problem is, however, that while in
QED the fluxes $f_T^{\gamma}(x,P^2)$ and $f_L^{\gamma}(x,P^2)$ are
known, we do not know how to calculate analogous fluxes of
transverse and longitudinal gluons inside hadrons. In fact it would
be sufficient to know the {\em relative} size of these fluxes, but
even this is not calculable. Nevertheless, guided by the situation
for photon fluxes, we can expect them to be similar, in particular
at low $x$. If this is assumed, the considerations of the preceding
subsection applied to gluons imply that the NLO gluonic
contribution to $F_2^{\mathrm p}(x,Q^2)$ comes from the sum of
contributions of transverse and longitudinal gluons. This in turn
means that the finite nonlogarithmic term in the gluonic
coefficient function $C^{(1)}_G(x)$ should be taken as
$-2+12x(1-x)$ rather than the usual $-2+6x(1-x)$.

\subsection {Virtuality dependence of initial conditions}
There are currently two approaches to introducing virtuality
dependence of PDF of the photon. Schuler--Sj\"{o}strand base their
parameterizations \cite{sas1} on the idea, suggested in
\cite{bjorken}, to use dispersion relations in $P^2$ written for
moments of $F_2^{\gamma}(x,P^2,Q^2)$. Their parameterizations do not
satisfy the same evolution equations
(\ref{Sigmaevolution}-\ref{NSevolution}) as those of the real
photon, but as the difference is formally of power correction type
this is no principal drawback.

In a different approach pursued in \cite{GRS,GRS2}, PDF of the
virtual photon are assumed to satisfy the same evolution equations
(\ref{Sigmaevolution}-\ref{NSevolution}) as those of the real
photon, and their virtuality dependence is introduced via the
virtuality dependence of the initial conditions at $Q^2=P^2$
\footnote{To retain the notation of \cite{GRS,GRS2} I use in this
subsection $Q^2$ instead of $M^2$ to denote the factorization scale.}.
This dependence is assumed to interpolate
between the standard boundary conditions for the real photon, which
in the GRV approach are defined at a very low scale
$Q^2=\mu^2<1$ GeV$^2$, and the predictions of perturbation theory
\begin{equation}
D(x,P^2,\tilde{P}^2)=\eta(P^2)D_{\mathrm {NP}}(x,\tilde{P}^2)+[1-\eta(P^2)]
D_{\mathrm {PT}}(x,\tilde{P}^2),~~ D=q,\overline{q},G,
\label{ini}
\end{equation}
where $\tilde{P}^2={\mathrm {max}}(P^2,\mu^2)$, $\eta(P^2)=
(1+P^2/m_{\rho}^2)^{-2}$ and
\begin{eqnarray}
D_{\mathrm {NP}}(x,\tilde{P}^2)& = & \kappa\left(2\pi\alpha/
f_{\rho}\right)\left\{
\begin{array}{lc}
f^{\pi}(x,P^2), & P^2>\mu^2\\
f^{\pi}(x,\mu^2), & 0\le P^2\le\mu^2
\end{array}\right.,
\label{iniNP} \\
q_{\mathrm {PT}}(x,\tilde{P}^2)& = &
\kappa\left(2\pi\alpha/f_{\rho}\right)
\left\{
\begin{array}{l}
0,~~{\mathrm {LO}}\\
2e_q^2\frac{\alpha}{2\pi}\left[(x^2+(1-x)^2)
\ln\frac{1}{x^2}-2+6x(1-x)\right],~~{\mathrm {NLO}}
\end{array}\right. \label{QPTini}\\
G_{\mathrm {PT}}(x,\tilde{P}^2) & = & 0,
\label{GPTini}
\end{eqnarray}
where $\kappa,f_{\rho},\mu^2,f^{\pi}(x,\mu^2)$ are given in
\cite{GRV2,GRV3}. The above boundary conditions are assumed to be
valid in the DIS$_{\gamma}$ factorization scheme (see next
subsection). The difference in (\ref{QPTini}) between the form of
boundary conditions on quark distribution functions at LO and NLO
follows again from misassignment of the coefficient function
$C^{0)}_{\gamma}(x)$ to the NLO. As argued in previous sections
this term is actually of the LO and therefore if taken into account
in boundary conditions, it must be present in both LO and NLO
analyses.

To include it would, however, go against the very essence of the
parton model interpretation of virtuality dependent PDF. To retain
clear physical interpretation of the factorization scale $Q$,
$q(x,P^2,Q^2)$ must vanish as the lower bound on the quark
virtuality $\tau^{\mathrm {min}}(x)\propto P^2$ approaches $Q^2$ from
below. This occurs explicitly in the expression (\ref{fullresult})
for all terms, whether logarithmic or not, that have parton model
interpretation. The term $\tau^{\mathrm {min}}/Q^2$, which guarantees
the vanishing of the nonlogarithmic part of (\ref{fullresult}) as
$\tau^{\mathrm {min}}\rightarrow Q^2$, is formally of the higher
twist and therefore invisible in \cite{bardeen} or any other
considerations within the leading twist approximation.

\subsection{Why DIS$_{\gamma}$ factorization scheme?}
For hadrons the LO relation between $F_2^{\gamma}$
quark distribution functions is the same in all FS and is also
identical to that of the parton model. DIS FS was introduced to
retain this simple relation also at the NLO, where different FS
correspond to different NLO splitting functions $P^{(1)}_{kl}(x)$
and hard scattering cross--sections (coefficient functions for
$F_2$). However, once the latter is chosen, the former is uniquely
determined (or vice versa). For proton the DIS FS amounts to
setting $C^{(1)}_q=0$.

For photon the situation is different. The DIS$_{\gamma}$ FS was
introduced for real photon to get rid of the troubling
$\ln(1-x)$ term in the LO expression (\ref{rightLO}) for
$C^{(0)}_{\gamma}(x,0,1)$. It implies setting
$D_{\gamma}=C^{(0)}_{\gamma}$ and $E_{\gamma}=0$ in
(\ref{redefinition}) and thus
modifies the inhomogeneous splitting functions $k^{(1)}$ and
$k^{(2)}$, but does not change the NLO homogeneous splitting functions
$P^{(1)}_{kl}$. As emphasized in Section 3 this redefinition of quark
distribution functions has in fact little to do with QCD.
\begin{figure}\centering
\epsfig{file=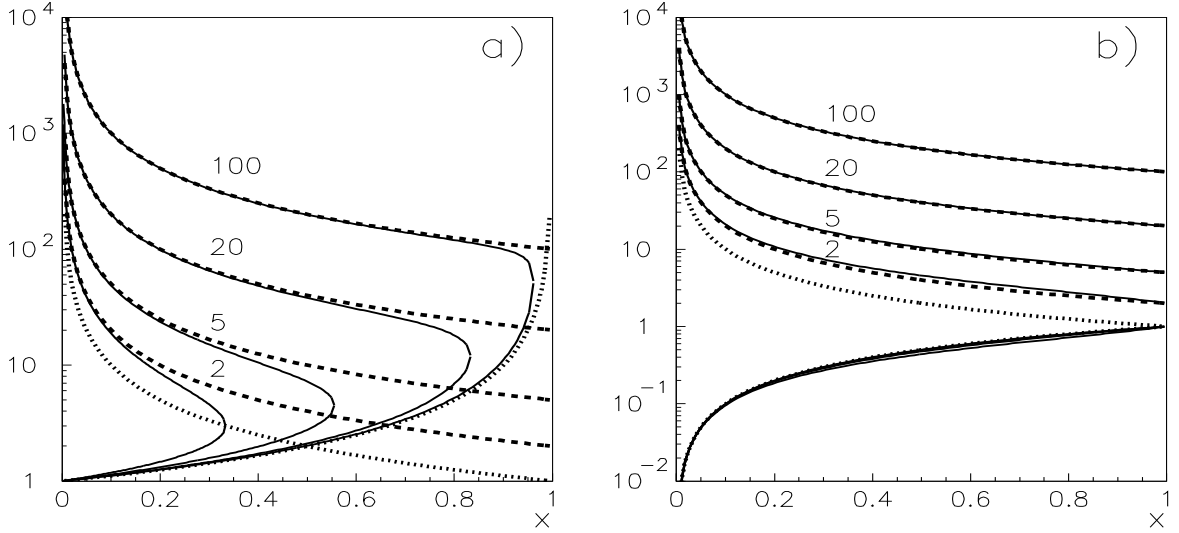,width=\textwidth,height=7cm}
\caption{Exact as well as approximate
bounds on the ratio $\tau/m_q^2$ for real photons and massive
quarks (a) and on the ratio $\tau/P^2$ for virtual photons coupled
to massless quarks (b). In both cases solid and dashed curves
correspond to exact and approximate bounds respectively, plotted
for four values of the ratio $Q^2/m^2$ in a) and $Q^2/P^2$ in b).}
\label{Figbounds}
\end{figure}
In parton model derivation of Section 4, the troubling term
$\ln(1-x)$ is a direct consequence of the fact that for real photon
parallel singularity associated with the splitting
$\gamma\rightarrow q\overline{q}$
is regulated by means of quark masses. This is natural in
QED, but as emphasized in subsection 4.2 not in QCD, where
nonperturbative properties of hadrons (with the exception of
pseudoscalar mesons) and the photon are expected to result from the
dynamics of confinement and we can therefore start from massless
quarks.

Note that the term $\ln(1-x)$ in $C^{(0)}_{\gamma}(x,0,1)$ causes
problems primarily because it appears there decoupled from the value
of the   quark mass $m_q$ with which it originally entered the
considerations in the expression (\ref{kinematics}) for
$\tau^{\mathrm min}$ and thus persists there even in the limit
$m_q\rightarrow 0$. It comes from the
lower bound on the quark virtuality $\tau$ in collinear kinematics
\begin{equation}
\tau_{\mathrm {min}}^{\mathrm {coll}}\equiv \frac{m_q^2}{1-x}\le \tau\le
\tau_{\mathrm {max}}^{\mathrm {coll}}\equiv \frac{Q^2}{x}.
\label{bounds}
\end{equation}
The ratio $Q^2(1-x)/xm_q^2$ of upper an lower integration limits in
(\ref{bounds}) equals unity (and thus the integral vanishes) at
$x_m\equiv 1/(1+m_q^2/Q^2)$, which for light quarks is very close
to $1$. For $x\rightarrow 1$ collinear kinematics is no
longer appropriate and should be replaced with the exact one
\begin{equation}
\frac{y}{2}\frac{1-\sqrt{1-z}}{x}
\le \frac{\tau}{m_q^2}\le \frac{y}{2}
\frac{1+\sqrt{1+z}}{x},
\label{exactbounds}
\end{equation}
where $y\equiv Q^2/m_q^2$ and $z\equiv 4yx/(1-x)$. These exact
bounds are shown, for several values of $y$, in Fig.
\ref{Figbounds}a by the solid curves, together with the dashed
ones, corresponding to the approximate bounds as given in
(\ref{bounds}). Also shown by the dotted curves are the functions
$1/x$ and $1/(1-x)$.

On the other hand when the term $\ln((1-x)/x)$ alone is added (as
part of the expression for $C^{(0)}_{\gamma}$) to the pointlike
solution of the inhomogeneous evolution equation with initial
conditions at $M_0\gg m_q^2$, the sum vanishes already at
$x_M\equiv 1/(1+M_0^2/Q^2)\ll x_m$ and becomes independent of $m_q$
down to $m_q=0$! However, it would be strange to retain in
$C_{\gamma}^{(0)}$ the term that follows directly from assuming
nonzero quark masses, but which when combined with the PDF is then
unrelated to their values. In the presence of confinement it seems
more appropriate to start from the very beginning with massless
quarks coupled to off--shell photons and then construct the limit
$P^2\rightarrow 0$. This approach is even more natural taking
into account that all current
data on the structure of the ``real'' photon actually come from
interactions of virtual photons, albeit with small virtuality. For
them the bounds on quark virtuality in collinear kinematics
\begin{equation}
xP^2\le \tau \le \frac{Q^2}{x}
\label{virtapp}
\end{equation}
are only slightly changed using exact kinematics
\begin{equation}
1+\frac{(1-x)y}{2}\frac{(1-\sqrt{1+z})}{x}
 \le \frac{\tau}{P^2} \le 1+\frac{(1-x)y}{2}\frac{(1+\sqrt{1+z})}{x}
\label{virtex}
\end{equation}
where now $y\equiv Q^2/P^2$ and $z\equiv 1/(1-x)-xy/(1-x)$. The
modification of (\ref{virtapp}) is so small that some of the solid
curves in Fig. \ref{Figbounds}b are indistinguishable from the
corresponding dashed one. Our suggestion is therefore to use also for
the real photon the expression
\begin{equation}
C^{(0)}_{\gamma}(x,0,1)=\left[x^2+(1-x)^2\right]
\ln\frac{1}{x^2}-2+8x(1-x),
\label{C0taky}
\end{equation}
derived for the transverse photon with virtuality $P^2\gg m_q^2$.
This term is different from (\ref{c0virtwrong}) normally
used for the virtual photon, as the latter corresponds to
$F_{2,T}^{\gamma}-F_{2,L}^{\gamma}/2$.

\section{Conclusions}
We have discussed the factorization mechanism for
$F_{2}^{\gamma}(x,Q^2)$ and pointed out the differences with
respect to the case of proton structure function
$F_{2}^{\mathrm p}(x,Q^2)$ which are due to the
presence of the inhomogeneous term in evolution equations for
$q^{\gamma}$. On the one hand, this term allows us to calculate the
asymptotic behaviour of $F_{2}^{\gamma}(x,Q^2)$ as $Q^2\rightarrow
\infty$ but on the other hand the same term also implies that a
complete NLO analysis of $F_{2}^{\gamma}(x,Q^2)$ requires the
inclusion of the ${\cal O}(\alpha_s^2)$ terms $k_{q}^{(2)},~k_G^{(2)}$
in the inhomogeneous splitting functions $k_{q},~k_G$, as well as of
the ${\cal O}(\alpha_s)$ term $C_{\gamma}^{(1)}$ term in photonic
coefficient
function $C_{\gamma}^{q}$. Unfortunately, none of them has been
included in the existing phenomenological analyses. There is no
problem to include the latter as the necessary calculations are
available, and we have shown that its numerical importance may be
quite large. On the other hand, $k_{q}^{(2)}$ and $k_G^{(2)}$ are
not known as their evaluation requires three loop QCD
calculation. Consequently, at the present
time a complete NLO analysis of $F_{2}^{\gamma}(x,Q^2)$ is
impossible to perform.

As far as the structure of the virtual photon is concerned, we have
discussed the question of what is actually measured in DIS on the
virtual photon and emphasized the role of the longitudinal photon
in these considerations.

We have also analyzed parton model interpretation
of the constant terms in $C_{\gamma}^{(0)}$ and discussed their
implications for the specification of initial conditions imposed on
the PDF of the virtual photon. We have argued that the presence of
colour confinement washes out the difference between the
regularization of mass singularities in the case of real and
virtual photons. This offers us a simple way of avoiding the
problems with the term $\ln(1-x)$ appearing in the
standard expression for $C_{\gamma}^{(0)}$ in the case of the real
photon.

\vspace*{0.5cm}
\noindent
\vspace*{0.5cm}
{\Large \bf Acknowledgments}

\noindent
I am grateful to W. van Neerven, E. Laenen and S. Larin for
correspondence concerning higher order QCD calculations of photonic
coefficient and splitting functions.

\end{document}